\documentclass[fleqn,twoside]{article}
\usepackage[headings]{espcrc2}
\usepackage[dvips]{epsfig}
\usepackage{here}

\readRCS $Id: espcrc2.tex,v 1.2 2004/02/24 11:22:11 spepping Exp $
\usepackage{graphicx}
\usepackage[figuresright]{rotating}

\newcommand{\AmS}{{\protect\the\textfont2
  A\kern-.1667em\lower.5ex\hbox{M}\kern-.125emS}}
\newcommand{\pt}{$p_{\rm T}$}
\newcommand{\fnref}[1]{$^{\ref{#1}}$}

\title{QCD studies with ATLAS at the LHC }

\author{V.\ A.\ Mitsou\address[IFIC]{Instituto de F\'{i}sica Corpuscular
(IFIC), CSIC –- Universitat de Val\`{e}ncia, \\
Edificio Institutos de Investigaci\'{o}n, Apartado de Correos 22085 \\
E-46071 Valencia, Spain}, on behalf of the ATLAS Collaboration}

\runtitle{QCD studies with ATLAS at the LHC } \runauthor{V.\ A.\ Mitsou}

\begin{document}

\begin{abstract}
The study of QCD processes at the LHC will serve two main goals. First, the
predictions of Quantum Chromodynamics will be tested and precision measurements
will be performed, allowing additional constraints to be established, and
providing measurements of the strong coupling constant. Second, QCD processes
represent a major part of the background to other Standard Model processes and
signals of new physics at the LHC and therefore need to be understood in depth.
An overview of various measurements of QCD-related processes to be performed at
the LHC is presented, based on final states containing high-\pt\ leptons,
photons and jets. Moreover, possible deviations from QCD predictions indicating
presence of new physics are discussed.

\vspace{1pc}
\end{abstract}

\maketitle

\section{INTRODUCTION}

\vspace{0.13cm}

\begin{figure}[H]
\vspace*{-0.5cm}
\centering\epsfig{width=0.83\linewidth,file=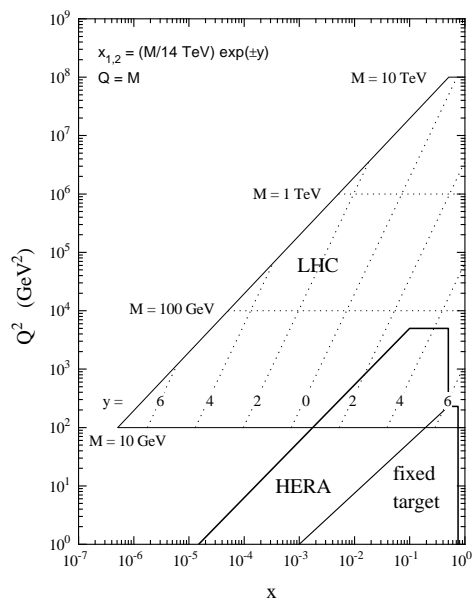,clip=}
\vspace*{-0.9cm}
\caption{Parton kinematics at the LHC in the $(x,Q^2)$ kinematic plane
for the production of a particle of mass $M$ at rapidity $y$
\protect\cite{stirling}.} \label{fig:LHC_xQ2}
\end{figure}

The Large Hadron Collider (LHC) is a proton-proton collider with a 14-TeV
centre-of-mass energy planned to be operated at a luminosity of $10^{34}~{\rm
cm^{-2}s^{-1}}$. This luminosity will result in large event samples for most
processes such as $\sim\!\!10^8$ leptonic W~decays, $\sim\!\!10^4$ photons with
$p_{\rm T}>500$~GeV and $\sim\!\!10^4$ jets with $p_{\rm T}>1$~TeV. With these
data, the theory of strong interactions will be precisely tested and detailed
measurements can be performed leading to new information on the parton
densities of the proton and possibly to constraints of the strong coupling
constant in yet unexplored regions. Figure~\ref{fig:LHC_xQ2} displays the
region in the $(x,Q^2)$ plane which will be covered by LHC in comparison with
HERA and fixed-target experiments.

\vspace{0.13cm}

This report presents an overview of QCD studies at the LHC, assuming the
expected performance of the ATLAS experiment, which is briefly described in
Section~\ref{sec:atlas}. First, issues related to minimum-bias events
(Section~\ref{sec:minbias}) and hard diffractive scattering
(Section~\ref{sec:diffr}) are presented. Next, the information to be deduced
from the measurements of jets (Section~\ref{sec:jets}) is described, followed
by a section on photon physics (Section~\ref{sec:photons}) and one concerning
the production of Drell-Yan pairs and heavy gauge bosons
(Section~\ref{sec:DY}). In Section~\ref{sec:heavy}, the production of heavy
flavours is discussed. Before concluding, the discovery of new physics, such as
quark compositeness, through the observation of deviations from QCD predictions
is addressed. More details on QCD studies in ATLAS can be found in
Ref.~\cite{TDR1}.

\begin{table*}[htb]
\caption{Basic performance characteristics of the ATLAS detector for the LHC.}
\label{table:atlas}
\newcommand{\m}{\hphantom{$-$}}
\newcommand{\cc}[1]{\multicolumn{1}{l}{#1}}
\renewcommand{\tabcolsep}{2.5pc} 
\renewcommand{\arraystretch}{1.2} 
\begin{tabular}{@{}ll}
\hline
Detector sub-system       & \cc{Performance} \\
\hline
Tracking (Si + transition radiation detector)
& $\sigma/p_{\rm T}\simeq5\cdot10^{-4}~p_{\rm T}{\rm (GeV)}+1\%$ \\
EM calorimeter (Pb + liquid Ar)
& $\sigma/E\simeq10\%/\sqrt{E{\rm (GeV)}}$ \\
Hadronic barrel calorimeter (steel + scintillator)
& $\sigma/E\simeq50\%/\sqrt{E{\rm (GeV)}}+3\%$  \\
Hadronic end-cap calorimeter (Cu/W + liquid Ar)
& $\sigma/E\simeq60\%/\sqrt{E{\rm (GeV)}}+3\%$  \\
Muon spectrometer (air toroidal magnet)
& $\sigma/p_{\rm T}\simeq10\%$ at $p_{\rm T}\!\sim\,$1~TeV \\
\hline
\end{tabular}\\[2pt]
\end{table*}

\section{THE ATLAS EXPERIMENT}\label{sec:atlas}

The ATLAS detector \cite{proposal} is a general-purpose experiment designed to
be sensitive to the various physics processes expected to take place at the
LHC. The design luminosity of $10^{34}~{\rm cm^{-2}s^{-1}}$ will allow for
high-statistics samples to be collected, resulting in measurements limited in
precision mostly by systematic uncertainties. Most of the QCD-related studies
will take place during the initial operation at a (low) luminosity of
$10^{33}~{\rm cm^{-2}s^{-1}}$, delivering a total of $10~{\rm fb^{-1}}$ per
year.

\begin{figure}[H]
\vspace*{-0.7cm}
\centering\epsfig{width=0.88\linewidth,file=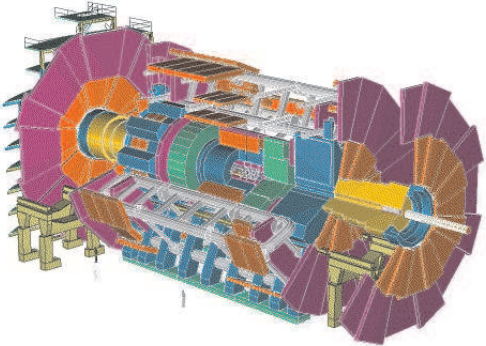}
\vspace*{-0.9cm}
\caption{Schematic view of the ATLAS detector.}
\label{fig:atlas}
\vspace*{-0.5cm}
\end{figure}

This hermetic apparatus provides pseudorapidity coverage of up to $|\eta|<2.5$
for tracking, extended to $|\eta|<5$ for calorimetry, while muons will be
accurately measured up to $|\eta|<2.7$. It comprises a high-performance tracker
consisting of silicon detectors and a transition radiation tracker in a 2-T
solenoidal magnetic field, a high-resolution electromagnetic calorimeter based
on lead/liquid argon, a hadronic calorimeter combining steel/scintillator and
Cu-W/liquid argon, and a large muon spectrometer embedded in an air-core
toroidal magnet. The basic performance parameters of these systems are given in
Table~\ref{table:atlas} and a schematic view of the ATLAS detector is shown in
Fig.~\ref{fig:atlas}.

\section{MINIMUM-BIAS INTERACTIONS}\label{sec:minbias}

Due to the high luminosity at the LHC, there will be up to an average of
25~inelastic collisions per bunch-crossing. In order to understand precisely
their contribution to the measured quantities for the hard scattering events of
interest, a detailed knowledge of the structure of the minimum-bias events is
required.

\begin{figure}[H]
\vspace*{-0.4cm}
\centering\epsfig{width=0.9\linewidth,file=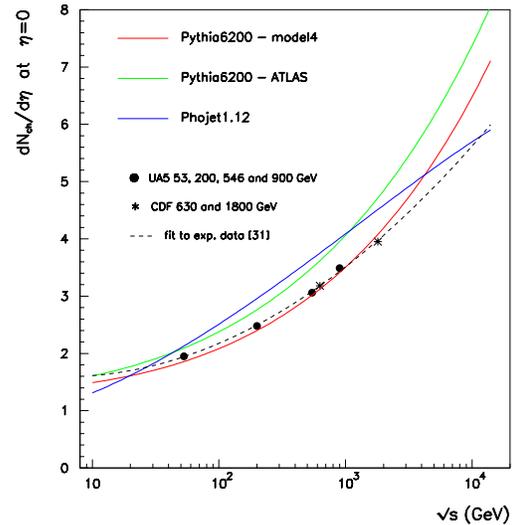}
\vspace*{-0.9cm}
\caption{Charged particle density ${\rm d}N_{\rm ch}/{\rm d}\eta$ at
$|\eta|=0$ as a function of $\sqrt{s}$ \protect\cite{minbias}. PYTHIA and
PHOJET predictions are compared with UA5 and CDF data.} \label{fig:min_bias}
\end{figure}
\vspace*{-0.5cm}

Several studies have been done performing comparisons between various Monte
Carlo simulation programs. In particular \cite{minbias}, PYTHIA6.200
\cite{PYTHIA} and PHOJET1.12 \cite{PHOJET} predictions are found to be in good
agreement with experimental data for energies below 800~GeV, as far as the
inelastic cross section is concerned, however they diverge at the level of
$\sim\!15\%$ at higher energies. Figure~\ref{fig:min_bias} displays the charged
particle density ${\rm d}N_{\rm ch}/{\rm d}\eta$ at $|\eta|=0$ as a function of
$\sqrt{s}$ per pp collision. The generators predictions are compared with
experimental data from UA5 \cite{UA5} and CDF \cite{CDF}. PYTHIA6.200 provides
a better description for energies up to $\sim\!3~{\rm TeV}$, nevertheless the
extrapolation of the fit-to-experimental-data to the LHC energy favours
PHOJET1.12. More studies are needed to develop accurate theoretical models for
minimum-bias events.

\section{HARD DIFFRACTIVE PROCESSES}\label{sec:diffr}

The understanding of diffractive phenomena, i.e.\ processes that arise from the
exchange of colour-neutral objects, has received revived attention in the last
few years due to the appearance of hard diffractive process, i.e.\ diffractive
processes in which a hard scatter takes place. They appear either as single
(\mbox{$\rm AB\rightarrow AX$} or \mbox{$\rm AB\rightarrow BX$}) or double
(\mbox{$\rm AB\rightarrow ABX$}) diffractive processes. Diffractive events are
characterized by the occurrence of large rapidity gaps and, in the case of
single-diffractive events, by the appearance of a leading hadron, i.e.\ a
hadron with a momentum close to the beam momentum separated from the
diffractive final state X.

The advantage of the LHC is in the production of diffractive final states with
large masses, allowing the probing of partonic structure with a variety of
different processes \cite{diffr}. A selection of events with two leading
protons or rapidity gaps on both sides of the detector transforms the
proton-proton collider into a Pomeron-Pomeron collider with variable beam
energy, where the maximal centre-of-mass energy ranges between the one of the
$\rm Sp\bar{p}S$ and the Tevatron collider.

An extension of the LHC detectors in the very forward region beyond $|\eta|=5$
(Roman Pot systems more than 100~m away from the interaction point) is planned
by the CMS \cite{CMS} (TOTEM \cite{TOTEM} detector) collaboration and is under
study by the ATLAS collaboration. It would increase the acceptance for charged
particles from inelastic interactions and provide tagging and measurements of
leading protons from elastic and diffractive interactions.

\section{JET SIGNATURES}\label{sec:jets}

Measurements of jets allow conclusions to be drawn on the hard scattering
process taking into account the evolution of the partonic system from the hard
scattering to the observed set of hadrons, i.e.\ the parton showering, the
fragmentation, the short-lived particle decays and the multiple interactions
\cite{jets}. Important systematic uncertainties (the statistical ones are small
at the LHC) are, e.g., the jet algorithm, the energy scale (especially at very
large $E_{\rm T}$) and the underlying event.

\begin{figure}[t]
\centering\epsfig{width=0.9\linewidth,file=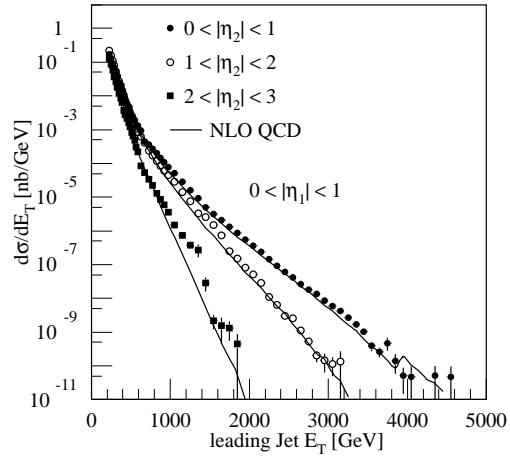}
\vspace{-0.9cm}
\caption{Di-jet cross section at hadron level with a leading jet with
$|\eta_1|<1$ for different ranges of the pseudorapidity of the second leading
jet as obtained from PYTHIA (points) and from a NLO Monte Carlo calculation
(solid line) \protect\cite{jets}.}\label{fig:dijet_tdr}
\vspace*{-0.5cm}
\end{figure}

The expected statistics for inclusive jet production at the LHC for an
integrated luminosity of $30~{\rm fb^{-1}}$ amounts to $4\cdot10^4$ events with
$E{\rm _T^{jet}>1~TeV}$, 3000 events with $E{\rm _T^{jet}>2~TeV}$ and
$\sim\!40$ events with $E{\rm _T^{jet}>3~TeV}$. Figure~\ref{fig:dijet_tdr}
shows the di-jet differential cross section for different jet rapidities and
for a minimal transverse energy of 180~GeV. The measurement of di-jet events
and their properties for different values of the minimal $E_{\rm T}$ and of the
two jet pseudorapidities can be used to constrain the parton densities inside
the proton in various kinematic regions of the $(x,Q^2)$ plane.

\begin{figure}[htb]
\vspace{-0.5cm}
\centering\epsfig{width=0.9\linewidth,file=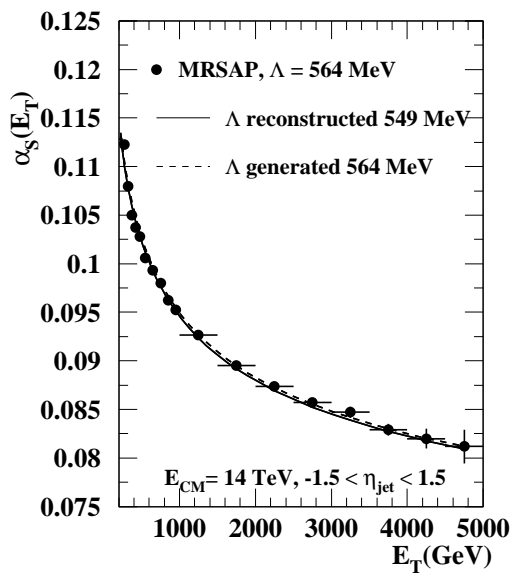}
\vspace*{-0.9cm}
\caption{Extracted values of $\alpha_s(E_T)$ (points), fitted with the evolution
function to determine $\Lambda^{(4)}_{\overline{\rm MS}}$ (solid line) and the
evolution used in the generation of the cross section (dashed line)
\protect\cite{alphas}. }\label{fig:alpha_s}
\end{figure}

A study \cite{alphas} was carried out by ATLAS to evaluate a method which
allows the extraction of the strong coupling constant, $\alpha_s$, from an
inclusive jet cross section measurement based on a parameterization of the
$\alpha_s$ dependence. The approach taken is to fit the jet cross section as a
function of $E_{\rm T}$ and $\alpha_s(E_{\rm T})$ in two ranges of
pseudorapidity and study its sensitivity to different sets of parton
distribution function (pdf). By inverting this fit, the determination of
$\alpha_s(E_{\rm T})$ is possible. An example of this analysis is shown in
Fig.~\ref{fig:alpha_s}, where both the generated and reconstructed values of
$\alpha_s(E_{\rm T})$ are plotted. The residual bias in terms of
$\alpha_s(E_{\rm T})$ is in average of the order of a few percent. The results
indicate that $\alpha_s$ can be determined with an overall precision of
$\sim\!10\%$, in which the contribution to the uncertainty from our knowledge
of pdf's is $\sim\!3\%$.

\section{PHOTON SIGNATURES}\label{sec:photons}

\begin{figure}[htb]
\vspace*{-0.4cm}
\centering\epsfig{width=0.9\linewidth,file=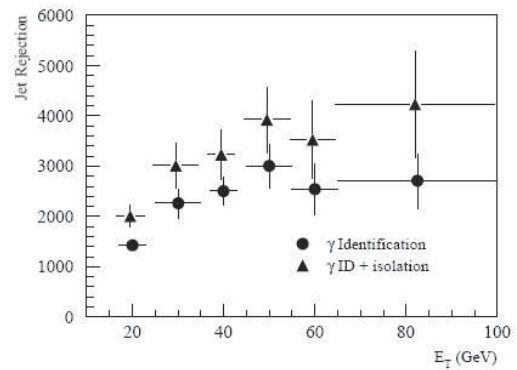}
\vspace*{-0.9cm}
\caption{Jet rejection after photon identification and isolation
cuts as a function of the jet $E_{\rm T}$ at low luminosity. The tuning was
done so as to achieve an 80\% efficiency for photons coming from Higgs decays
\protect\cite{isolation}.} \label{fig:isol}
\vspace*{-0.5cm}
\end{figure}

Direct photon measurements can provide important constraints on parton
distributions, especially on the gluon distribution in the proton. The
advantage of photon measurements is the better energy determination in
comparison to jet measurements. In the case of photons, however, the
experimental background due to jets containing a leading $\pi^0$ has to be well
understood. Photon identification in ATLAS is based on the shower shapes in the
calorimeters, conversion reconstruction and a track veto. However, $\gamma/{\rm
jet}$ separation power can be enhanced if isolation criteria \cite{isolation}
are applied around the shower, i.e.\ no significant hadronic activity is
allowed in a cone around the photon direction. Figure~\ref{fig:isol} shows the
dependence of the jet rejection on the transverse energy at low luminosity with
and without isolation cuts. For an 80\% photon efficiency for photons from
${\rm H}\rightarrow\gamma\gamma$ decays, a jet rejection of 1440 (880) is
achieved at low (design) luminosity.

\begin{figure}[htb]
\vspace*{-0.4cm}
\centering\epsfig{width=0.9\linewidth,file=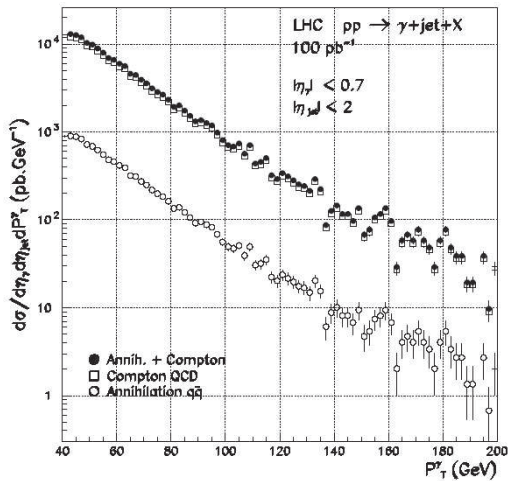}
\vspace*{-0.9cm}
\caption{Direct photon production with an opposite-side jet as a function of
$p_{\rm T}^{\gamma}$ obtained from PYTHIA and the CTEQ4L structure function
\protect\cite{gluon}.} \label{fig:gluon1}
\vspace*{-0.5cm}
\end{figure}

The direct production of photons can provide sensitivity to the gluon density
in the proton via the QCD Compton process, $\rm qg\rightarrow \gamma q$
\cite{gluon}. Figure~\ref{fig:gluon1} displays the differential cross section
for direct photon production with an opposite-side jet
\mbox{($150^\circ<\Delta\phi_{\rm \gamma-jet}<210^\circ$)} for the LHC energy.
The QCD Compton process clearly dominates over the annihilation graph, $\rm
q\bar{q}\rightarrow \gamma g$. The CTEQ4L \cite{CTEQ} set is used for the
parameterization of the proton, while other sets such as CTEQ4HJ, MRS(A)
\cite{MRS} and GRV94HO \cite{GRV} give similar results within 10\%.

\begin{figure}[htb]
\centering\epsfig{width=0.9\linewidth,file=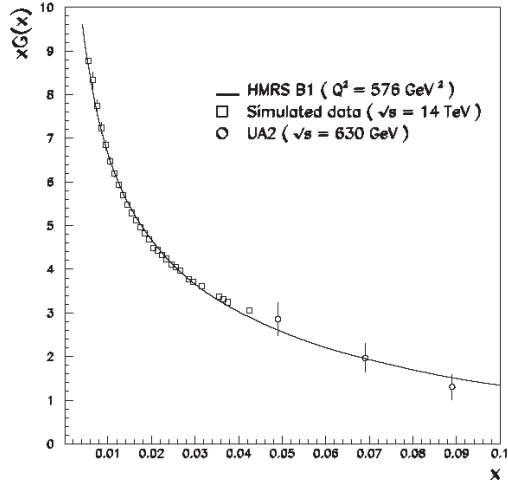}
\vspace*{-0.9cm}
\caption{Comparison between theoretical prediction, ATLAS simulated data
and data from the UA2 experiment \protect\cite{gluon}.} \label{fig:gluon2}
\vspace*{-0.5cm}
\end{figure}

The method used to extract the gluon structure function, $xG(x)$, is based on
fitting the $\rm\gamma\!-\!jet$ cross section with the theoretical prediction
at LO. In this example, the MRS(A) set is used to parameterize the quarks pdf's
and the CTEQ set is adopted for $xG(x)$. The results for the gluon density in
the proton are shown in Fig.~\ref{fig:gluon2} in comparison with UA2 \cite{UA2}
data and the HMRS~B1 \cite{HMRS} theoretical prediction, with which they are in
good agreement. The gluon density can be determined in the range $0.005<x<0.04$
and $440~{\rm GeV^2}<Q^2<2\cdot10^4~{\rm GeV^2}$ with a systematic error of
10--20\%, which can be reduced if NLO calculation is used for the cross
section.

\section{LEPTONIC FINAL STATES}\label{sec:DY}

The measurement of Drell-Yan lepton pair production and the production of W and
Z bosons (with a leptonic decay to electrons or muons) will allow constraints
to be set on the quark and anti-quark densities of the proton at a scale given
by the invariant mass of the lepton pair or by the W/Z boson mass over a wide
range in Bjorken $x$ \cite{TDR1}. QCD effects enter the cross section of these
processes only in the initial state, making thus the measurements less
uncertain.

For an integrated luminosity of $30~{\rm fb^{-1}}$, about $10^5$ events are
expected which contain a W boson with $p{\rm _T^W}>400~{\rm GeV}$ decaying to
electron or muon and a neutrino. The expected number for Z production is
smaller by an order of magnitude. Given the large statistics and the clean
signatures expected for W and Z production, there is a possibility to use these
processes for a precise determination (at a level of few percent) of the
parton-parton luminosity. Furthermore, the production of gauge boson pairs will
provide the opportunity to study electroweak parameters such as the triple
gauge boson couplings.

\section{HEAVY QUARKS}\label{sec:heavy}

The production of heavy quarks, due to the quark mass involved, provides an
important process for the study of perturbative QCD and of non-perturbative
aspects. The total cross section for charm production is 7.8~mb, the one for
beauty production is 0.5~mb, while the top-pair production cross section is
about 0.8~nb.

Due to a large beauty production cross section and a selective trigger, ATLAS
can reconstruct large samples of exclusive B-hadron decays \cite{bphys}.
Statistically dominant are the exclusive channels with $\rm
J/\psi\rightarrow\mu\mu$, which will allow measurements up to $p_{\rm
T}\sim100~{\rm GeV}$ with negligible statistical errors ($\sim\!2000$ events
with $p_{\rm T}>100~{\rm GeV}$).

\begin{figure}[htb]
\vspace*{-0.4cm}
\centering\epsfig{width=0.9\linewidth,file=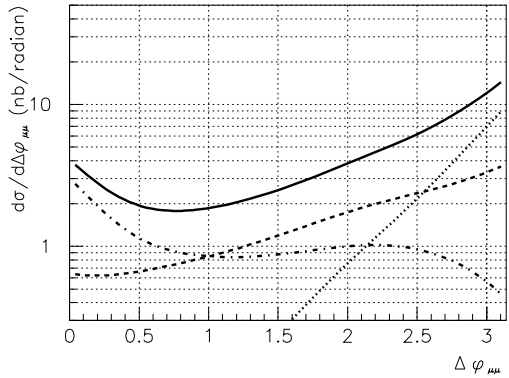}
\vspace*{-0.9cm}
\caption{Different mechanisms contributing to azimuthal $\mu\mu$ correlations
at the LHC: flavour excitation ${\rm gb\rightarrow gb}$ (dash-dotted line),
gluon-gluon fusion ${\rm gg\rightarrow b\bar{b}}$ (dotted line), gluon-gluon
scattering followed by gluon splitting ${\rm gg\rightarrow gg}$ with ${\rm
g\rightarrow b\bar{b}}$ (dashed line), and the sum of all contributions (solid
line) \protect\cite{bphys}.}\label{fig:bphys}
\vspace*{-0.5cm}
\end{figure}

Correlations between b and $\rm\bar{b}$ quarks, which were difficult to be
studied in previous experiments due to limited statistics, will be investigated
in detail. In Fig.~\ref{fig:bphys} the expected azimuthal correlation
$\Delta\phi_{\mu\mu}$ between the muons from the b and $\rm\bar{b}$ decays is
shown, which provides information on the $\rm b\!-\!\bar{b}$ correlation. The
domain of back-to-back kinematics, $\Delta\phi_{\mu\mu}\sim\pi$, is mostly
populated by LO QCD contribution. In contrast, the effects of higher orders are
more pronounced in the $\Delta\phi_{\mu\mu}\sim0$ region, which is free of the
LO contribution.

Possible deviations from QCD expectations (like new $s$-channel resonances)
should give characteristic signatures in the invariant mass of the $\rm
t\bar{t}$ pair, hence the study of such kinematical distributions is important.
Besides that, the extraction of the $\rm H/A\rightarrow t\bar{t}$ decay
requires a precise knowledge on the invariant mass distribution, which
represents a background process for this channel. Within perturbative QCD the
total cross section for top pair production in higher-order corrections is
sensitive to the top mass and the scale uncertainty. An error of 1\% on the top
mass corresponds to an error of 5\% in the total $\rm t\bar{t}$ cross section
\cite{TDR1}. Assuming a top quark mass of 175~GeV, the total cross section for
$\rm t\bar{t}$ production is 803~pb at NLO and 833~pb for NLO including the NLL
resummation.

\section{QUARK COMPOSITENESS}\label{sec:compo}

The observation of deviations from QCD predictions of jet rates will reveal new
physics such as quark compositeness \cite{compo} or the existence of new
particles. Measuring the inclusive jet cross section and studying the di-jet
mass spectrum and angular distributions are essential tests of QCD.

In the study \cite{TDR2} carried out by ATLAS, the quark substructure, which
manifests itself over a compositeness scale $\Lambda$, is simulated with PYTHIA
\cite{PYTHIA}. The data simulated in the framework of the Standard Model (SM)
are compared with those obtained assuming quark compositeness.

\begin{figure}[htb]
\vspace*{-0.4cm}
\centering\epsfig{width=0.9\linewidth,file=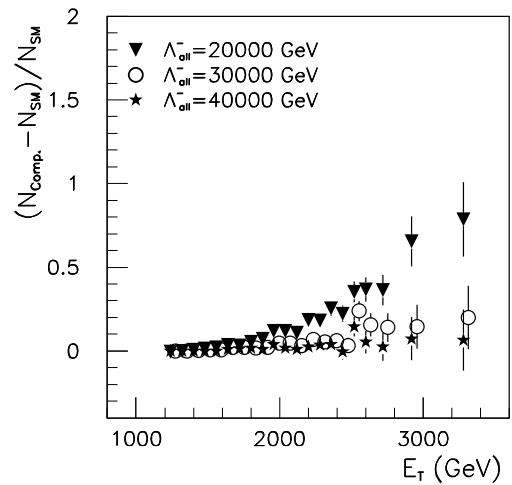}
\vspace*{-0.9cm}
\caption{Difference between the SM prediction and the effect of compositeness
on the jet $E_{\rm T}$ distribution, normalized to the SM rate for an
integrated luminosity of 300~${\rm pb^{-1}}$
\protect\cite{TDR2}.}\label{fig:fig23_24}
\end{figure}

Figure~\ref{fig:fig23_24} shows the effect of compositeness (all quarks are
assumed to be composite) on the inclusive jet energy spectrum, normalized to
the SM prediction. The deviation is significant only for large values of
$E_{\rm T}$. However such a signal may be faked either by the uncertainties in
the pdf's (equivalent to a signal for $\Lambda=15~{\rm TeV}$) or by the
non-linear response of the hadron calorimeter. The pdf's are expected to be
further constrained before and after LHC starts running (by LHC as well),
eliminating thus this background source. If, on the other hand, the calorimeter
non-linearity was understood at the 1.5\% level, a sensitivity at 95\%
confidence level (CL) up to $\Lambda$ equal to 25~(40)~TeV for 30~(300)~$\rm
fb^{-1}$ should be feasible.

\begin{figure}[htb]
\vspace*{-0.4cm}
\centering\epsfig{width=0.9\linewidth,file=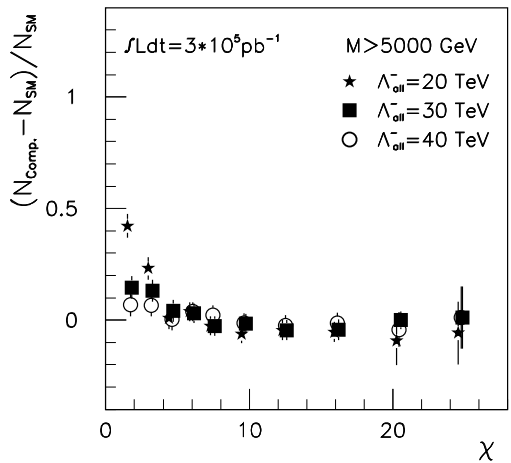}
\vspace*{-0.9cm}
\caption{Di-jet angular distribution for di-jet mass above 5000~GeV for an
integrated luminosity of 300~${\rm pb^{-1}}$
\protect\cite{TDR2}.}\label{fig:fig23_20}
\vspace*{-0.5cm}
\end{figure}

The angular distribution of the jets is more sensitive to compositeness signals
than the jet transverse energy spectrum and less susceptible to calorimeter
non-linearities. The analysis is made in terms of the angular variable
$\chi\equiv\exp|\eta_1-\eta_2|$, where $\eta_{1,2}$ are the pseudorapidities of
the two leading jets. Figure~\ref{fig:fig23_20} shows the deviation of the
di-jet angular distribution from the SM predictions for invariant di-jet mass
larger than 5~TeV. It is clear that quark compositeness leads to an enhancement
in the distribution at low values of $\chi$ when compared with the SM
predictions.

In conclusion, the study demonstrates that the high-mass di-jet angular
distribution has an excellent discovery capability for quark compositeness. One
month of LHC operation at low luminosity allows discovering quark substructure
if the constituent interaction constant is 14~TeV. An integrated luminosity of
$\rm300~fb^{-1}$ is needed to reach a 95\% CL limit of 40~TeV.

\section{SUMMARY}\label{sec:summary}

A variety of QCD-related processes can be studied with the ATLAS detector at
the LHC. These measurements are of importance as a study of Quantum
Chromodynamics, accessing a new kinematic regime at the highest energy
accessible in a laboratory. A precise knowledge and understanding of QCD
processes is also essential for the studies of the Higgs boson(s) and searches
for new physics beyond the Standard Model, where QCD represents a large part of
the background.

Candidate signatures to provide constraints on the quark and anti-quark
distributions are the production of W and Z bosons via the Drell-Yan process as
well as lepton pair production in general. The production of direct photons,
jets, beauty and top quarks can be used to acquire information on the gluon
density in the proton.

The precise measurement of the inclusive jet cross section will allow
constraints to be set on the strong coupling constant with an uncertainty of
$\sim\!10\%$. Furthermore, a possible discrepancy between the observed and the
theoretically predicted cross section may indicate the discovery of new physics
like quark compositeness.

The LHC will extend the kinematic range to larger values of $Q^2$, the hard
scale of the partonic scale, reaching scales of the order of $\rm 1~TeV^2$. The
fraction of the proton momentum attributed to a parton, $x$, will access values
below $10^{-5}$ with an energy scale above $\rm100~GeV^2$.

\section*{Acknowledgements}

This work has been performed within the ATLAS Collaboration and is the result
of collaboration-wide efforts. I would like to thank the organizers of the {\em
$\it 11^{th}$ International QCD Conference} for the stimulating environment
they offered during the event and M.~Dobbs for valuable comments on this
report. The author acknowledges support by the EU funding under the RTN
contract: HPRN-CT-2002-00292, {\em Probe for New Physics}.

\end{document}